\documentclass[12pt]{article}

\usepackage{amsmath}
\usepackage{amsfonts}
\usepackage{amssymb}
\usepackage{graphicx}
\usepackage{xcolor}

\usepackage{hyperref}
\usepackage[font=footnotesize,labelfont=bf]{caption}
\newcommand{\w}{\wedge}

\newcommand{\nn}{\nonumber}

\newcommand{\omeg}[2]{ \omega^{#1}_{\ #2}}

\begin{document}

\begin{titlepage}
\title{Accelerating anisotropic cosmologies in Brans-Dicke gravity coupled to a  mass-varying vector field}
\author{\"{O}zg\"{u}r Akarsu\footnote{E.mail: oakarsu@ku.edu.tr}, Tekin Dereli\footnote{E.mail: tdereli@ku.edu.tr},
Neslihan Oflaz\footnote{E.mail: noflaz@ku.edu.tr}
\\
\small{ Department of Physics, Ko\c{c} University}\\ {\small 34450 Sar{\i}yer, {\.I}stanbul, Turkey}}

\date { }

\maketitle

\begin{abstract}

\noindent The field equations of Brans-Dicke gravity  coupled to a mass-varying vector field are derived.  Anisotropic cosmological solutions with a  locally rotationally symmetric Bianchi type I metric and  time-dependent scalar and electric vector fields are studied. A particular class of exact solutions for which all the variable parameters have a power-law time dependence is given. The universe  expands with a constant expansion anisotropy within this class of solutions. We show that the accelerating expansion is driven by the scalar field and the electric vector field can be interpreted as an anisotropic dark matter source.

\noindent \small{\\
\textbf{Keywords:} Brans-Dicke scalar $\cdot$ mass-varying vector field $\cdot$ anisotropic universe models $\cdot$ accelerating expansion $\cdot$ isotropization }

\end{abstract}

\end{titlepage}

\section{Introduction}

The most successful cosmological model describing the observed features of the universe so far is arguably the $\Lambda$CDM (cold dark matter) model \cite{Padmanabhan03}. It has three fundamental assumptions: (i) the universe at large scales can be described by the spatially flat, isotropic and homogeneous Robertson-Walker (RW) metric (relying on the inflation scenario \cite{Linde08}), (ii) the constituents of the universe can be described by a positive cosmological constant $\Lambda$ together with distributions of CDM and baryonic matter and (iii) general relativity (GR) is the true theory of gravity. The latest data from the Planck cosmic microwave background (CMB) experiment, whose major goal is to test $\Lambda$CDM model to high precision and identify areas of tension, shows a remarkable consistency with the predictions of the base $\Lambda$CDM model \cite{PlanckXVI,PlanckXXIII,PlanckXXVI}. However, it also confirms the previously found anomalies in the large-scale CMB data from the WMAP experiment \cite{WMAP7an} that might be ascribed to the RW metric assumption \cite{PlanckXXIII,PlanckXXVI} and reveals a number of intriguing features of the data that might be ascribed to the $\Lambda$ assumption \cite{PlanckXVI}.

In the last decade we have not only accumulated data converging on the fact that $\Lambda$CDM model is the simplest successful model but also data questioning the fundamental assumptions of this model. There is a large literature that argues that CMB multipole alignments, QSO polarization alignment and large scale bulk flows all prefer approximately the same direction in the sky. Preference of a similar direction has recently been shown in the CMB maximum temperature asymmetry axis \cite{Antonio13} and in the direction dependence of the acceleration of the universe \cite{Antoniou10,Cai12,Zhao13}. The Planck experiment also concludes that the most significant large-scale anomalies in the statistical isotropy of the CMB temperature, namely the quadrupole-octopole alignment, hemi-spherical asymmetry and etc., represent real features of the CMB sky \cite{PlanckXXIII}. These observed anomalies individually may not be conclusive but taken together they bring the isotropy assumption into question. For instance, it was showed that the CMB quadrupole problem can be solved without giving rise to a new problem when the large-scale spatial geometry of the universe is allowed to be ellipsoidal with eccentricity at decoupling of order $10^{-2}$ \cite{Campanelli06}.

 It is well-known that the generic inflationary model relies on scalar fields and predicts an almost completely isotropic universe as a result of 60 e-folds increase of the size of the universe during the inflationary epoch \cite{Linde08}. Hence, if it is true that the space is actually anisotropic, then one should either introduce an inflationary scenario in which a small anisotropy could survive at the end of the inflation or introduce a mechanism that can anisotropize the universe slightly after the inflation took place. The most obvious way of altering the isotropization process/inducing anisotropy is maybe to introduce anisotropic sources. Analyses of cosmological evolution with known matter sources that possess small anisotropic pressures; electric/magnetic fields, spatial curvature anisotropies, anti-symmetric axion fields, simple topological defects and etc. are given in \cite{Barrow97}. However, because such sources should have been dominated by dust and then dark energy (DE) since the decoupling, in regard of the possibly slightly anisotropic geometry of the universe, the possibility of anisotropic models of inflaton and/or DE sources (e.g., vector field models of these sources) comes into question.

The possibility of an anisotropic inflation driven by a vector field was first suggested in 1989 \cite{Ford89} but the idea began to attract interest only recently \cite{Golovnev08,Koivisto08,Kanno08,Watanabe09,Kanno10,Bartolo13a,Bartolo13b}. However, it should be noted that inflation models where accelerated expansion is driven by a vector field usually suffer from ghost instabilities \cite{Himmetoglu09a,Himmetoglu09b}. Alternatively, generic scalar field inflation may be kept as it is and some anisotropy can be induced in relatively recent times relying on a DE source that yields an anisotropic equation of state (EoS), e.g. a vector field, and hence accelerates the universe anisotropically \cite{Koivisto08,Koivisto06,Mota08,Rodrigues08,Battye09,Akarsu10a,Akarsu10b,Campanelli11,Thorsrud12}. The possibility of anisotropic DE and anisotropic  acceleration in the present time universe are subjects of current observational studies as well \cite{Campanelli11,Mota07,Appleby10,Appleby13}.

It is found in the Planck experiment that the CMB data alone is compatible with $\Lambda$ assumption of $\Lambda$CDM but a DE component with a time varying EoS parameter is favored when the astrophysical data is also taken into account in the analysis \cite{PlanckXVI}. The study of such DE sources with a time-varying EoS parameter has begun right after the discovery of the current acceleration of expansion of the universe \cite{Riess98,Perlmutter99}, regarded as alternatives to $\Lambda$ within the context of GR particularly due to the theoretical problems associated with the value of $\Lambda$ \cite{Padmanabhan03,Zeldovich,Weinberg89}. Such models of DE were mostly based on the existence of scalar fields that can mimic $\Lambda$ under appropriate conditions \cite{Copeland06,Bamba12}. However, the scalar field models of DE also face problems similar to those of $\Lambda$ and require further ad hoc assumptions for their introduction. An alternative to the presence of a DE source, on the other hand, is to consider modified theories of gravity such as the well-known Brans-Dicke theory of gravity \cite{Brans-Dicke,Clifton12}. In fact  many modified theories of gravity realized by augmenting GR with at least one or more extra degrees of freedom are expected to give rise in general to late time cosmic acceleration. It is not as obvious as it is in the case of anisotropic sources, but the modification of GR may also modify the isotropization process (See e.g. Ref.\cite{Barrow06,Barrow10}). Accordingly if the accelerated expansion of the current universe is going to be  attributed to a modification of GR rather than DE, then the isotropization process must also be modified in some particular way.

In this paper, we couple a mass-varying vector field to the Brans-Dicke scalar-tensor theory of gravity where the extra scalar degree of freedom may induce both an accelerated expansion and also modify the isotropization in a particular way. In order to discuss cosmological solutions, we start with a spatially homogeneous and flat but not necessarily isotropic LRS Bianchi I space-time metric. We also introduce a time-dependent, homogeneous electric field vector to induce anisotropy. The varying mass of the vector field is fixed as a function of the scalar field for consistency of our field equations and is constrained to real positive values since an imaginary (tachyonic) mass for the vector field leads to a ghost instability (see \cite{Himmetoglu09a,Himmetoglu09b} and references therein for details). The accelerated expansion of the universe in our model turns out not to be driven by the vector field contrary to other vector field models in the literature, but rather by the scalar field. Our vector field behaves more like a dark matter source with an anisotropic EoS that is responsible of a slight, persistent anisotropy of the acceleration.

\newpage

\section{Field equations}

We derive the non-minimally coupled Brans-Dicke-vector field equations from the infinitesimal variations of the action density
\begin{equation}
\mathcal{L}[e,\phi,A] = \frac{\phi}{2} \mathcal{R} \ast 1 - \frac{\omega}{2\phi}{\rm d}\phi \w \ast {\rm d}\phi - \frac{1}{2} F \w \ast F - \frac{m^2[\phi]}{2} A \w \ast A ,
\end{equation}
where $A$ is the vector potential 1-form of Proca with $F = {\rm d}A$. $\omega$ is the Brans-Dicke parameter and $m[\phi]$ is the variable vector boson mass that is given as a function of the scalar field $\phi$.  We  find it convenient to re-express the action density in terms of a new scalar field $\alpha^2 = \phi$ and consider the variations of
\begin{equation}
\label{action}
\mathcal{L} = \frac{\alpha^2}{2}\mathcal{ R} \ast 1 - 2\omega {\rm d}\alpha \w \ast {\rm d}\alpha - \frac{1}{2} F \w \ast F - \frac{m^2[\alpha]}{2} A \w \ast A,
\end{equation}
subject to the constraint that the space-time torsion vanishes. The
co-frame variations of (\ref{action}) give the Einstein field equations
\begin{eqnarray}
\label{ee}
 - \frac{\alpha^2}{2}R^{bc} \w \ast ( e_a \w e_b \w e_c)=  D(\iota_a\ast d \alpha^2) + 4 \omega \tau_a[\alpha] + \tau_a[F] + m^2[\alpha]  \tau_a[A] ,
\end{eqnarray}
where the stress-energy 3-forms on the right hand side are given by
\begin{eqnarray}
\label{tau}
 \tau_c[\alpha]& = & \frac{1}{2} \left( \iota_a {\rm d}\alpha \w \ast {\rm d}\alpha + {\rm d}\alpha \w \iota_a \ast {\rm d}\alpha   \right),
 \\
 \tau_c[F]&=&\frac{1}{2}\left( \iota_a F \w \ast F - F \w \iota_a \ast F \right),
\\
\tau_a[A]&=&\frac{1}{2}\left( \iota_a A \w \ast A + A \w \iota_a \ast A \right) ,
\end{eqnarray}
respectively. Here $ \iota_a$'s denote the interior product operators dual to the co-frames, that is, $\iota_a (e^b) = \delta^{a}_{\;b}$.
We obtain from the $\alpha$-variations of (\ref{action}),
\begin{equation}
\label{eqn:rev1}
\alpha^2 \mathcal{R}\ast 1 + 4\omega \alpha {\rm d}\ast {\rm d} \alpha-\frac{\alpha}{2} \frac{{\rm d}m^2}{{\rm d}\alpha}A \w \ast A=0.
\end{equation}
We then trace the Einstein field equations (\ref{ee}) and use it together with (\ref{eqn:rev1}) to eliminate the curvature scalar term. Thus we find the scalar field equation
\begin{equation}
\label{alpha}
(2\omega+3) {\rm d} \ast {\rm d}\alpha^2  + \left( m^2 - \frac{\alpha}{2}\frac{{\rm d} m^2}{{\rm d}\alpha}\right) A \w \ast A = 0.
\end{equation}
While the variations of (\ref{action}) with respect to $A$ yield the vector field equation
\begin{equation}
\label{A}
{\rm d}\ast F + m^2 \ast A = 0.
\end{equation}

\section{The cosmological ansatz}

We start by assuming a time-dependent scalar field
\begin{equation}
\alpha=\alpha(t)
 \end{equation}
 and a time-dependent, spatially homogeneous potential 1-form
\begin{equation}
\label{eqn:A}
A= \beta(t) {\rm d}z.
\end{equation}
Since  $F = \dot{\beta} {\rm d}t \wedge {\rm d}z$ is an electric field along the $z$-direction, the rotational symmetry of the space is broken. Accordingly, we  introduce two metric scale factors; $a(t)$ in the $z$-direction that is pointed by the vector field and $b(t)$ for the transverse $x$- and $y$-directions and work with a locally rotationally symmetric (LRS) Bianchi type-I metric tensor
\begin{equation}
g = -{\rm d}t \otimes {\rm d}t + a^2(t) {\rm d}z \otimes {\rm d}z + b^2(t) \left ( {\rm d}x \otimes {\rm d}x + {\rm d}y \otimes {\rm d}y \right ).
\end{equation}
With the choice of an orthonormal co-frame as
\begin{equation}
e^0 = {\rm d}t,\ e^1= a(t) {\rm d}z,\ e^2 = b(t) {\rm d}x,\ e^3 = b(t) {\rm d}y;
\end{equation}
the Levi-Civita connection 1-forms $\{\omega^{a}_{\; b}\}$ can be solved from the first Cartan structure equations
\begin{equation}
 {\rm d} e^a + \omega^a_{\ b} \wedge e^b = 0.
\end{equation}
We find
\begin{equation}
\omeg{0}{\; 1}=\frac{\dot{a}}{a} e^1,\quad \omeg{0}{\;2}=\frac{\dot{b}}{b} e^2, \quad \omeg{0}{\;3}=\frac{\dot{b}}{b} e^3, \quad \omeg{1}{\;2} = \omeg{2}{\;3}=\omeg{1}{\;3}=0,
\end{equation}
in terms of which the curvature 2-forms are calculated from the second Cartan structure equations
\begin{equation}
R^{a}_{\; b} = {\rm d} \omeg{a}{b} + \omeg{a}{c} \w \omeg{c}{ b}.
\end{equation}
The resulting curvature 2-forms are
\begin{eqnarray}
R^{0}_{\;1}=\frac{\ddot{a}}{a} e^0 \w e^1,\quad R^{0}_{\;2}=\frac{\ddot{b}}{b} e^0 \w e^2,\quad R^{0}_{\;3}=\frac{\ddot{b}}{b} e^0 \w e^3,\\
\nn
R^{1}_{\;2}=\frac{\dot{a}\dot{b}}{ab} e^1 \w e^2,\quad R^{2}_{\;3}=\frac{\dot{b}^2}{b^2} e^2 \w e^3,\quad R^{1}_{\;3}=\frac{\dot{a}\dot{b}}{ab} e^1 \w e^3 ,
\end{eqnarray}
and the corresponding Ricci scalar is
\begin{equation}
\mathcal{R} = 2\left(\frac{ \ddot{a}}{a}+2 \frac{\ddot{b}}{b}+ 2\frac{\dot{a}\dot{b}}{a b} +  \frac{\dot{b}^2}{b^2}\right).
\end{equation}

\medskip

\noindent The set of coupled o.d.e's that describes our cosmological model can now be obtained by substituting the above expressions into the field equations (\ref{ee}), (\ref{alpha}) and (\ref{A}). We get the following  vector field equation
\begin{equation}
\label{mass}
\frac{\ddot{\beta}}{\beta} + \left( -\frac{\dot{a}}{a} + 2 \frac{\dot{b}}{b}\right)\frac{\dot{\beta}}{\beta} + m^2[\alpha] = 0 ,
\end{equation}
and the scalar field equation
\begin{equation}
\label{dildiff}
(2\omega+ 3) \left[\frac{\ddot{\alpha}}{\alpha} + \left ( \frac{\dot{\alpha}}{\alpha} \right )^2 + \left (\frac{\dot{a}}{a}+2\frac{\dot{b}}{b} \right )\frac{\dot{\alpha}}{\alpha}\right] = \frac{\beta^2}{2 a^2 \alpha^2} \left( m^2 - \frac{\alpha}{2}\frac{{\rm d}{m^2}}{{\rm d}\alpha} \right ) .
\end{equation}
Three other independent equations are provided by the Einstein field equations:
\begin{eqnarray}
\label{eediff1}
2\frac{\dot{a}\dot{b}}{ab}+ \frac{\dot{b}^2}{b^2} - 2\omega\frac{\dot{\alpha}^2}{\alpha^2 } + 2 \frac{\dot{\alpha}}{\alpha}\left(\frac{\dot{a}}{a}+2\frac{\dot{b}}{b}\right) = \frac{\beta^2}{2 \alpha^2 a^2} \left(  \frac{\dot{\beta}^2}{\beta^2} + m^2 \right),
\\
\label{eediff2}
2\frac{\ddot{b}}{b}+ \frac{\dot{b}^2}{b^2} + (2\omega+2)\frac{\dot{\alpha}^2}{\alpha^2 } + 2 \frac{\ddot{\alpha}}{\alpha}+ 4\frac{\dot{\alpha}}{\alpha}\frac{\dot{b}}{b} = \frac{\beta^2}{2 \alpha^2 a^2} \left( \frac{\dot{\beta}^2}{\beta^2}- m^2 \right),
\\
\label{eediff3}
\frac{\ddot{a}}{a}+\frac{\ddot{b}}{b}+\frac{\dot{a}\dot{b}}{ab} + (2\omega+2)\frac{\dot{\alpha}^2}{\alpha^2 } + 2 \frac{\ddot{\alpha}}{\alpha}+2 \frac{\dot{\alpha}}{\alpha}\left(\frac{\dot{a}}{a}+\frac{\dot{b}}{b}\right) = \frac{\beta^2}{2 \alpha^2 a^2} \left( -\frac{\dot{\beta}^2}{\beta^2} + m^2 \right) .
\end{eqnarray}
We thus end up with a system of five ordinary differential equations (\ref{mass})-(\ref{eediff3}) to be solved. It can be checked that this system is fully determined, yet they are far too complicated for us to be able to write down a general analytic solution. However, we can construct a family of exact solutions assuming a power-law behavior in cosmic time $t$ for all the variables involved. In the next section, we first discuss some physical parameters related with anisotropic cosmological models that can be handled without making use of any explicit solution. In the section that follows we present the exact power-law solution and then discuss the corresponding cosmological model in detail.

\section{Physical parameters of the model}
\label{sec:general}

It would be convenient at this stage to introduce some cosmological parameters that we shall later use. Namely, we define the average scale factor $v$, the average Hubble parameter $H$ and the deceleration parameter of the volumetric expansion, respectively,  as follows:
\begin{equation}
v=\left (ab^2 \right )^{\frac{1}{3}},\quad H=\frac{1}{3}\left(\frac{\dot{a}}{a}+2\frac{\dot{b}}{b}\right), \quad q= -\frac{v\ddot{v}}{\dot{v}^2}= -1 + \frac{{\rm d}}{{\rm d}t}\left(\frac{1}{H}\right). \label{23}
\end{equation}
In a similar way, the directional Hubble parameters and the directional deceleration parameters along the $x$-, $y$- and $z$-axes will be given:
\begin{equation}
H_{x}=H_{y}=\frac{\dot{b}}{b},\quad H_{z}=\frac{\dot{a}}{a}, \quad  q_{x}=q_{y}=-1+\frac{{\rm d}}{{\rm d}t}\left(\frac{1}{H_{x}}\right),\quad q_{z}=-1+\frac{{\rm d}}{{\rm d}t}\left(\frac{1}{H_{z}}\right).
\end{equation}
The deceleration parameters are the key parameters among the others, because for any expanding scale factor (namely $v$, $a$ or $b$), the negative values of the corresponding deceleration parameter imply acceleration, positive values deceleration and the special values $-1$ and $0$ correspond either to exponential expansion or the constant-rate expansion, respectively.
Two further cosmological parameters relevant to the discussion of anisotropic cosmological models are the shear scalar
\begin{equation}
\label{eqn:aniso1}
\sigma^{2}=\frac{1}{2}\sum_{i=1}^{3}\left(H_{i}-H\right)^{2} ,
\end{equation}
and the expansion anisotropy parameter
\begin{equation}
\label{eqn:aniso2} \Delta=\frac{1}{3}\sum_{i=1}^{3}\left(\frac{H_{i}-H}{H}\right)^{2}
\end{equation}
where the sums are on  $(1,2,3)=(x,y,z)$. $\Delta$ is a measure of the deviation from isotropic expansion; the universe expands isotropically for $\Delta=0$. The time-evolution of $\Delta$ is crucial for deciding whether the universe approaches isotropy at some stage or not. In particular, the spatial section of the metric approaches isotropy for $v\rightarrow\infty$ and $\Delta\rightarrow 0$ as $t\rightarrow \infty$ \cite{CollinsHawking}. In order to determine the shear scalar and the expansion anisotropy, we consider the difference between the expansion rates of the $z$- and the $x$- or $y$-axes. Then subtracting (\ref{eediff3}) from (\ref{eediff2}) and re-organizing the resultant equation we get
\begin{equation}
\frac{{\rm d}}{{\rm d}t}(H_{z}-H_{x})+3H(H_{z}-H_{x})+2\frac{\dot{\alpha}}{\alpha}(H_{z}-H_{x})=-\frac{1}{\alpha^2 a^2}\left(\dot{\beta}^2-m^2\beta^2\right).
\end{equation}
The integration of this equation determines the difference between the expansion rates along the $z$- and the $x$- or $y$-axes as follows:
\begin{equation}
H_{z}-H_{x}=\frac{1}{\alpha^2 ab^2} \left[\lambda-\int{\frac{b^2}{a}\,(\dot{\beta}^2-m^2 \,\beta^2) } {\rm d} t \right] \label{26}
\end{equation}
where $\lambda$ is an integration constant. Using (\ref{26}) and the average Hubble parameter defined in (\ref{23}), we obtain the shear scalar
\begin{equation}
\sigma^2=\frac{1}{3\alpha^4 a^2b^4} \left[\lambda-\int{\frac{b^2}{a}\,(\dot{\beta}^2-m^2 \,\beta^2) } {\rm d} t \right]^{2},
\end{equation}
and the expansion anisotropy
\begin{equation}
 \Delta=\frac{2}{9H^2\alpha^4 a^2b^4} \left[\lambda-\int{\frac{b^2}{a}\,(\dot{\beta}^2-m^2 \,\beta^2) } {\rm d} t \right]^{2}
\end{equation}
as well.
We note that the difference between the expansion rates along the $x$- and $z$- axes and the square root of the shear scalar are both inversely proportional to the volume of the universe and  the square of the scalar field. In the case of a model based on GR alone, the square root of the shear scalar would have been simply inversely proportional to the volume of the universe.
Here the electric vector field also contributes to the shear scalar, hence to the isotropization history of the universe in a non-trivial way through the integral term in the above expressions.
For our massive vector field, given the energy density and the pressures along the $x$-, $y$- and $z$-axes
\begin{equation}
\rho = \frac{1}{2a^2}(\dot{\beta}^2+m^2\beta^2) , \quad
p_x = p_y =-p_z = \frac{1}{2a^2}(\dot{\beta}^2-m^2\beta^2),
\end{equation}
we immediately observe that the corresponding EoS parameters along the $x$-, $y$-  and $z$-axes are
\begin{equation}
w_{x}=w_{y}=-w_{z}= \left (\frac{\dot{\beta}^2}{\beta^2}-m^2 \right )/\left ({\frac{\dot{\beta}^2}{\beta^2}+m^2}\right ).
\end{equation}
It is known that tachyonic mass ($m^2<0$) for the vector field means that the longitudinal mode due to the mass is a ghost and leads to an instability in the model \cite{Himmetoglu09a,Himmetoglu09b}. In our model, on the other hand, we assume that $m$ is a real function (i.e., $m^2>0$) to evade such a ghost instability. The mass term $m^2\beta^2$ contributes positively to the energy density $\rho$ and directional EoS parameters can only take values between $-1$ and $1$. Some limiting cases are as follows: (i) $\frac{\dot{\beta}^2}{\beta^2}>>m^2 \rightarrow w_{x}=w_{y}=-w_{z}\sim 1$, the EoS is similar to those of the (static) electric/magnetic fields and of  cosmic strings; (ii) $\frac{\dot{\beta}^2}{\beta^2}<<m^2 \rightarrow w_{x}=w_{y}=-w_{z}\sim -1$, the EoS is similar to those of cosmic domain walls; (iii) $\frac{\dot{\beta}^2}{\beta^2}\sim m^2 \rightarrow w_{x}=w_{y}=-w_{z}\sim 0$ the EoS is similar to that of a dust. Therefore, the electric vector field in our model may not only describe anisotropic EoS different than that in the electromagnetic theory but also provides a dynamical anisotropic EoS parameter in a certain range and can even become isotropic in some cases.


\section{The exact power-law solution}
\label{sec:powersol}

In this section, we present a particular solution of the system for which all the variable parameters are of the power-law form with respect to the cosmic time $t$, and then discuss the properties of this solution. Let
\begin{eqnarray}
a:= a_1 t^{p_a},\quad b:= b_1 t^{p_b},\quad \alpha:= \alpha_1 t^{-s},\quad \beta:= \beta_1 t^{-u}\quad\textnormal{and}\quad m:= m_1 t^r
\label{eqn:powers}
\end{eqnarray}
where $a_1$, $b_1$, $\alpha_1$, $\beta_1$, $m_1$ are the values of the variables at time $t=1$ and $p_a$, $p_b$, $s$, $u$, $r$ are the powers of $t$ to be determined. As a side remark we note that under these assumptions the mass of the vector field should be related to the scalar field with a power-law relation as
\begin{equation}
m[\alpha] = m_{1} \left(\frac{\alpha}{\alpha_{1}}\right)^{-\frac{r}{s}}.
\end{equation}
Substituting (\ref{eqn:powers}) into the system of  differential equations (\ref{mass})-(\ref{eediff3}), we immediately see that a consistent solution of the system requires
\begin{eqnarray}
r=-1,\quad p_a= s-u,\quad p_b= \frac{1}{2}+\frac{s}{2}+\frac{m_1^2}{2u},
\end{eqnarray}
making both sides of the vector field equation (\ref{mass}) vanish identically  and leading to the expressions
\begin{eqnarray}
\label{eqn:ara}
a= a_1 t^{s-u},\quad b= b_1 t^{\frac{1}{2}+\frac{s}{2}+\frac{m_1^2}{2u}},\quad \alpha = \alpha_1 t^{-s},\quad \beta= \beta_1 t^{-u}, \quad m= m_1 t^{-1}. \label{35}
\end{eqnarray}
 Now substituting these into the remaining equations (\ref{dildiff})-(\ref{eediff3}), we get a set of four algebraic relations:
\begin{eqnarray}
4(s-1) u - (11+8\omega) s^2 - 2s +1-4m_1^2  -2m_1^2(s-1) u^{-1} + m_1^4 u^{-2}& =& 2 \kappa^2 (u^2 + m_1^2), \nonumber \\
(11+8\omega) s^2 + 2s -1 -2m_1^2(s-1) u^{-1} + 3m_1^4 u^{-2} &=& 2 \kappa^2 (u^2 - m_1^2),
\nonumber \\
4u^2 - 2(s-1)u+(11+8\omega) s^2 + 2s -1-2m_1^2 +m_1^4 u^{-2}&=& 2 \kappa^2 (m_1^2-u^2),
 \nonumber \\
2s^2\left(2\omega+ 3\right) (u^2 - m_1^2)&=& \kappa^2 m_1^2 u (s-1),
\nonumber
\end{eqnarray}
where we set  $\kappa^2 =\frac{\beta_1^2}{a_1^2\alpha_{1}^2}$. Solution to this algebraic system parametrized in terms of $u$ and $m_1$, is given by
\begin{equation}
\label{kappa}
s =\frac{2u^4 - u^3 -m_{1}^2 (u^2-2u)+2m_{1}^4 }{u(-u^2+2m_{1}^2)}
\end{equation}
provided
\begin{equation}
\label{BDw}
\kappa^2=\frac{-2u^2 + 2m_{1}^2}{u^2 - 2m_{1}^2} \quad  \textnormal{and}
\quad \omega= -\frac{3}{2}+\frac{1}{2}\frac{m_{1}^2 u^2 (2m_{1}^4-m_{1}^2 u^2+2u^4)}{(2m_{1}^4-(u^2-2u)m_{1}^2+2u^4-u^3)^2} .
\end{equation}
 However, $u$ and $m_{1}$ cannot take arbitrary values: $\beta_{1}\neq 0$, $a_{1}$ and $\alpha_1$ are real parameters so that the positivity of $\kappa^2>0$ imposes a restriction on the possible values of the real parameters $u$ and $m_1>0$ according to
\begin{eqnarray}
\label{ineq}
m_1 < |u| <\sqrt{2}m_1.
\end{eqnarray}
This restriction on the parameter values is important since it in turn implies a constraint on the dynamics of the model as we shall see below.

\medskip

\section{Discussion}

\noindent We first of all summarize our power-law solution by explicitly giving
 the metric
\begin{equation}
g = -dt^2 + \frac{\beta_1^2}{\kappa^2 \alpha_1^2} t^{2s -2u} dz^2 + t^{1+s + {m_1^2}/u} (dx^2 + dy^2 )
\end{equation}
and the scalar and vector fields
\begin{equation}
\alpha = \alpha_1 t^{-s}  , \quad A = \beta_1 t^{-u} dz
\end{equation}
where
$s =\frac{2u^4 - u^3 -m_{1}^2 (u^2-2u)+2m_{1}^4 }{u(-u^2+2m_{1}^2)}$.
We have set $m[\alpha] = m_{1} (\frac{\alpha}{\alpha_{1}})^{-\frac{1}{s}}$.

It can be quickly verified that  the average Hubble  and the deceleration parameters of the universe turn out to be
\begin{eqnarray}
H &=& \frac{5u^4-3u^3 -5m_1^2u^2 +6m_1^4 u + 6m_1^4}{3u(2m_1^2-u^2)}\,t^{-1},
\label{eqn:Hpower}
\\
\label{q}
q &=& -1+\frac{3u(2m_1^2-u^2)}{5u^4-3u^3 -5m_1^2u^2 +6m_1^2 u + 6m_1^4}.
\end{eqnarray}
 It follows from (\ref{eqn:Hpower}) that the condition for getting an expanding universe $Ht>0$ is satisfied only in case the vector potential $\beta$ decreases as $t$ increases, i.e. only for $u>0$. This further reduces the inequality (\ref{ineq}) as follows:
\begin{eqnarray}
\label{ineq2}
m_1 < u <\sqrt{2}m_1.
\end{eqnarray}
The conditions $u>0$ and $m_{1}>0$ put the deceleration parameter in the range $-1<q<0$, which means that an expanding universe is necessarily be  accelerating in our model. Using the inequalities in (\ref{ineq2}), we find that the deceleration parameter reaches its minimum value
\begin{eqnarray}
q_{\rm min}\rightarrow -1\quad \textnormal{as}\quad  u\rightarrow \sqrt{2} m_1
\end{eqnarray}
and its maximum value
\begin{eqnarray}
q_{\rm max}\rightarrow -\frac{2m_1}{2m_1+1}\quad \textnormal{as}\quad u\rightarrow m_1.
\end{eqnarray}
We note that the larger $m_{1}$ becomes, the narrower is the range for the allowed values of the deceleration parameter. Namely, one may check that $q_{\rm max}\rightarrow q_{\rm min}\rightarrow -1$ for $m_1 >> 1$ so that only the exponential expansion is allowed in this limit.


In an anisotropic universe model we should also check  the expansion rates along different spatial axes. We obtain the directional Hubble parameters as
\begin{eqnarray}
\nonumber
H_x=H_y = & \frac{u^4- u^3-m_1^2 u^2+2m_1^2 u + 2m_1^4}{u(2m_1^2-u^2)}\, t^{-1},
\\
H_z = &\frac{3u^4- u^3-3m_1^2 u^2+2m_1^2 u+2m_1^4 }{u(2m_1^2-u^2)}\,t^{-1}.
\end{eqnarray}
It is easily verified that the universe expands in all directions since both $H_x$ and $H_z$ are positive for $u>0$ and $m_1>0$. The difference between the directional Hubble parameters will be\footnote{The expression that follows is consistent with (\ref{26}) provided the integration constant $\lambda =0$.}
\begin{equation}
\label{eqn:HminusH}
H_z-H_x=-\frac{2m_1^2-2u^2}{2m_1^2-u^2}\, \frac{u}{t}=\kappa^2\frac{u}{t},
\end{equation}
and hence the expansion rate of the universe along the $z$-axis is always greater than the expansion rate along the $x$- and $y$-axes (since $u>0$). On the other hand, the expansion rates along  different axes approach each other inversely in time $t$ and become identical as $t\rightarrow\infty$. Accordingly the shear scalar should be decreasing with the square of the cosmic time $t$ as follows:
\begin{eqnarray}
\sigma^2 = \frac{4}{3} u^2 \left(\frac{u^2-m_{1}^{2}}{u^2-2m_{1}^{2}}\right)^2 t^{-2}.
\end{eqnarray}
On the other hand, since the average Hubble parameter also decreases inversely  with the cosmic time $t$, the decay of the shear scalar leads neither to isotropization nor to anisotropization. In fact we have a  constant expansion anisotropy
\begin{equation}
\label{eqn:powerAn}
\Delta=\left(\frac{2\sqrt{2}u^2(u^2-m_1^2)}{5u^4-3u^3 -5m_1^2u^2 +6m_1^2 u + 6m_1^4}\right)^2
\end{equation}
 that takes values in the range $0<\Delta<\frac{1}{8}$ such that $\Delta\rightarrow 0$ as $u\rightarrow m_1$ while $\Delta\rightarrow \frac{1}{8}$ as $u\rightarrow \sqrt{2}m_1$.



In an accelerating anisotropic universe, not only the anisotropy of the expansion rate but also the anisotropy of the deceleration parameter is of interest. We obtain the directional deceleration parameters as follows:
\begin{eqnarray}
\nonumber
q_{x}=q_{y}= &-1+ \frac{u(2m_1^2-u^2)}{u^4- u^3-m_1^2 u^2+2m_1^2 u + 2m_1^4},
\\
q_{z}= &-1+\frac{u(2m_1^2-u^2)}{3u^4- u^3-3m_1^2 u^2+2m_1^2 u+2m_1^4 }.
\end{eqnarray}
The level of the anisotropy of the deceleration parameter can be quantified using the normalized difference according to
\begin{eqnarray}
\frac{\Delta q}{\bar{q}}=2 \frac{q_{z}-q_{x}}{q_{z}+q_{x}}={\frac {-2u^{3} ( 2m_1^{2}-u^{2}) (m_1^{2}-u^{2})}{4m_1^{8}+3u^{8}-2u^{7}+11m_1^{4} u^{4}-8m_1^{6}{u}^{2}-6m_1^{2} u^{6}+6m_1^{2}{u}^{5}-6m_1^{4}u^{3}+4m_1^{6}u}} \nonumber
\end{eqnarray}
which again takes a constant value.




Regarding the dynamics of the scalar and vector fields,  we first note that the Brans-Dicke coupling parameter $\omega$ can only take negative values of the order of $-1$ and exhibits some interesting limits:
\begin{eqnarray}
u\rightarrow m_1 \Longrightarrow \omega\rightarrow -\frac{3}{2}\;\frac{8m_1^2+6m_1+1}{9m_1^2+6m_1+1} \Longrightarrow
\begin{cases}
\omega\rightarrow -\frac{3}{2}\quad \textnormal{as}\quad m_1 \rightarrow 0 \\
\omega\rightarrow -\frac{4}{3}\quad \textnormal{as}\quad m_1 \rightarrow \infty
\end{cases}
\end{eqnarray}
and
\begin{eqnarray}
u\rightarrow \sqrt{2}m_1 \Longrightarrow \omega\rightarrow -\frac{11}{8}.
\end{eqnarray}
The above limiting values of $\omega$ are remarkable:  $-\frac{3}{2}<\omega<-\frac{4}{3}$ is the range for which  accelerated power-law expansion  exists within  Brans-Dicke gravity in the presence of dust \cite{Weinberg72}. We also find it interesting to note that the negative values  of $\omega$ of order of $-1$ can be motivated from the string theories. For instance, the case $\omega=-1$ appears in the low energy limit of effective string field theories and the case $\omega=-\frac{4}{3}$ appears in four-dimensional 0-brane space-times ($d = 1$) in  string models with $(d-1)$-branes \cite{Duff95,Lidsey00}.


Next we consider the energy density and the directional pressures of the vector field given by
\begin{equation}
\rho=\frac{\beta_{1}^2}{2a_{1}^2}(u^2+m_{1}^2)\,t^{-2s-2} \quad , \quad
p_{x}=p_{y}=-p_{z}=\frac{\beta_{1}^2}{2a_{1}^2}(u^2-m_{1}^2)\,t^{-2s-2}.
\end{equation}
Then the directional EoS parameters for the vector field will be constant:
\begin{equation}
\label{eqn:EOS}
w_{x}=w_{y}=-w_{z}=\frac{u^2-m_1^2}{u^2+m_1^2}.
\end{equation}
Using the allowed range for the values of $u$ with respect to $m_{1}$ (\ref{ineq2}), we find that the EoS can get values in the range
\begin{equation}
0<w_{x}=w_{y}=-w_{z} <\frac{1}{3}.
\end{equation}
It is interesting to note that for the choice $m_{1} \sim u$, the above distribution would be indistinguishable from a dust but with a small anisotropic pressure in contrast to the conventional dust fluid that is described with an isotropic EoS parameter $w=0$. Hence, in the present case, the vector field may be regarded as a dark matter fluid yielding a slightly anisotropic pressure. Moreover, defining the average EoS by $\tilde{w}=(w_x+w_y+w_z)/3$, we see that it only takes positive values $0<\tilde{w}<\frac{1}{9}$. This is another  indication that the vector field in our model cannot be interpreted as a possible DE source since a negative EoS parameter is characteristic of such sources.


Before we conclude, we wish to consider some observationally meaningful cosmological parameter values and evaluate other physically relevant quantities to check if our model gives consistent results or not.
Using the mean deceleration parameter value as $q=-0.7$ in (\ref{q}) and the
anisotropy of the expansion as $\Delta=10^{-4}$  in (\ref{eqn:powerAn}) \cite{Campanelli11},
we find a solution with $m_1 = 1.119$ and $u=1.136$ that satisfies (\ref{ineq2}). Using these two values in (\ref{eqn:EOS}), we find the directional EoS parameters of the vector field to be given by $w_{x}=w_{y}=-w_{z}=0.015$ and the average EoS  $\tilde{w}=0.005$. We note that such a fluid is indistinguishable from dust/CDM but yields a slightly anisotropic pressure. We find the values of the directional deceleration parameters as $q_{x}=q_{y}=-0.698$ along the $x$- and $y$-axes and $q_{z}=-0.704$ along the $z$-axis. Accordingly, the difference between the directional deceleration parameters along the $z$- and $x$- or $y$-axes will be $q_{z}-q_{x}=-0.006$. These imply a level of the anisotropy of the deceleration parameter  $\frac{\Delta q}{\bar{q}}=0.018$ that corresponds to a \% 1.8 level of difference. Finally  the Brans-Dicke coupling parameter value is determined to be $\omega=-1.399$.


In the present paper, we considered Brans-Dicke gravity coupled to a mass-varying vector field with the mass given as a real function of the scalar field. We derived the variational  field equations and looked for anisotropic cosmological solutions. Assuming a Bianchi type I metric and time-dependent, homogeneous scalar and electric vector fields, we obtained the reduced system of equations. We were able to construct, with a particular mass function,  a family of exact  solutions for which all the variables have power-law time-dependence. We discussed the physical predictions of such solutions in detail.  We showed in particular it is the scalar field which essentially drives the acceleration and that the vector field in our model should not be interpreted as a DE source but rather as a CDM source and that the interplay between the scalar and electric vector fields  would give rise to a non-trivial isotropization history of the universe.
Finally, we think that a wider investigation of our model possibly based on numerical analysis might be worth pursuing.

\bigskip
\begin{center}
\textbf{Acknowledgments}
\end{center}
\"{O}.A. acknowledges the postdoctoral research scholarship he is receiving from The Scientific and Technological Research Council of Turkey (T\"{U}B{\.I}TAK-B{\.I}DEB 2218). N.O. is supported in part by a Scholarship from The Scientific and Technological Research Council of Turkey (T\"{U}B{\.I}TAK). We all appreciate the support from Ko\c{c} University.

\end{document}